\documentclass[aps,pra,onecolumn,showpacs,amsmath,amssymb,floatfix,footinbib,superscriptaddress,24pt]{revtex4-1}
\usepackage{amsmath,natbib,graphicx,subfigure,amssymb,graphics,amsmath,mathrsfs,CJK,color}
\usepackage{multirow,fancyhdr,color,bm,tabularx,psfrag,dcolumn}
\usepackage[colorlinks=true,linkcolor=blue,urlcolor=blue,citecolor=blue]{hyperref}
\usepackage{tikz,xcolor,hyperref}
\definecolor{lime}{HTML}{A6CE39}
\DeclareRobustCommand{\orcidicon}{%
 \begin{tikzpicture}
 \draw[lime, fill=lime] (0,0)
    circle [radius=0.16]
    node[white] {{\fontfamily{qag}\selectfont \tiny ID}};\draw[white, fill=white] (-0.0625,0.095)
    circle [radius=0.007];
 \end{tikzpicture}
\hspace{-2mm}}
\foreach \x in {A, ..., Z}
{\expandafter\xdef\csname orcid\x\endcsname{\noexpand\href{https://orcid.org/\csname orcidauthor\x\endcsname}{\noexpand\orcidicon}}}

\begin{document}

\title{\large Differentiation of correlated fluctuations in site energy on excitation energy transfer in photosynthetic light-harvesting complexes}

\author{Lu-Xin Xu }
\affiliation{Center for Quantum Materials and Computational Condensed Matter Physics, Faculty of Science, Kunming University \\of Science and Technology, Kunming, 650500, PR China}
\affiliation{Department of Physics, Faculty of Science, Kunming University of Science and Technology, Kunming, 650500, PR China}

\author{Shun-Cai Zhao\textsuperscript{\orcidA{}}}
\email[Corresponding author: ]{zsczhao@126.com}
\affiliation{Center for Quantum Materials and Computational Condensed Matter Physics, Faculty of Science, Kunming University \\of Science and Technology, Kunming, 650500, PR China}
\affiliation{Department of Physics, Faculty of Science, Kunming University of Science and Technology, Kunming, 650500, PR China}

\author{Sheng-Nan Zhu }
\affiliation{Center for Quantum Materials and Computational Condensed Matter Physics, Faculty of Science, Kunming University \\of Science and Technology, Kunming, 650500, PR China}
\affiliation{Department of Physics, Faculty of Science, Kunming University of Science and Technology, Kunming, 650500, PR China}

\author{Lin-Jie Chen }
\affiliation{Center for Quantum Materials and Computational Condensed Matter Physics, Faculty of Science, Kunming University \\of Science and Technology, Kunming, 650500, PR China}
\affiliation{Department of Physics, Faculty of Science, Kunming University of Science and Technology, Kunming, 650500, PR China}
\date{\today}
\begin{abstract}
One of the promising approaches to revealing the photosynthetic efficiency of close to one unit is to investigate the quantum regime of excitation energy transfer (EET). The majority of studies, however, have concluded that different pigment molecules contribute equally to EET, rather than differently. We investigate the roles of different site-energies in EET by evaluating the correlated fluctuations of site-energies in two adjacent pigment molecules (namely Site 1 and Site 2), and we attempt to demonstrate different site roles in EET with the j-V characteristics and power via a photosynthetic quantum heat engine (QHE) model rather than an actual photosynthetic protein. The results show that fluctuations at Site 1 (the pigment molecule absorbing solar photons) provide ascending and then descending EET. At Site 2, the EET is reduced through the use of correlated fluctuation increments (the pigment molecule acting as the charge-transfer excited state). Furthermore, when investigating the correlated fluctuations at Site 2, the different gap differences of the output terminal play a positive role in EET, but a sharply decreasing EET process is also achieved with less correlated fluctuations at Site 2 compared to those at Site 1.The findings show that different pigment molecules contribute differently to EET. The significance of this work is that it not only clarifies the roles of different pigment molecules in EET, but it also deepens our understanding of the fundamental physics of EET as it transports through the molecular chain in photosynthetic light-harvesting complexes. Furthermore, the results are appropriate to the EET in organic semiconductors, photovoltaic devices, and quantum networks, when these systems couple to the environment of photons via the vibrational motion of sites in the molecular chain.
\begin{description}
\item[PACs]{42.50.-p, 32.80.Qk, 42.50.Gy}
\item[Keywords]{Quantum fluctuation; excitation energy transfer; two adjacent pigment molecules}
\end{description}
\end{abstract}

\maketitle
\section{Introduction}
Energy transport is a crucial and fundamental issue in many areas of physics, chemistry and biology \cite{2006How,2009Some,2008A,2010Noise,2010Decoherence,2010Effect2,2012Dephasing,RevModPhys.89.015001, 2017Environmental,2020Theoretical,10.1140/epjp/s13360-021-02009-3}, and transport efficiency in photosynthetic light-harvesting complexes and molecular crystals has received a lot of attention\cite{2005Exciton,2007Expansion,2008A,2009Some,2010Noise,2010Decoherence,2015A,2016Modeling}. The near-unity quantum efficiency of excitation energy transfer (EET) among pigments in a noisy environment\cite{2010Noise,Chain3377} inspires one to figure out its physical mechanism in order to illustrate design principles for artificial light-harvesting\cite{2010Photocatalytic,Meyer2011Chemist}.
Among the numerous theoretical models\cite{1997Exciton,2006How,2010Noise,2017Environmental,10.1140/epjp/s13360-021-02009-3,2020Theoretical,lizhao2021},the EET in the molecular chain was regarded as an open quantum system with excitations produced by the vibrational motion of sites in a quantum network. Given the complexities of photosynthetic systems, a few simplifying assumptions\cite{2006How,2008Environment,2009On} were made to facilitate calculations. Some of the results, such as energy transport\cite{2010Noise,2017Environmental,2020Theoretical} augmented by environmental noise, are obtained by employing the weak coupling and the Markov approximation\cite{2010Quantum1,2010Coherently,2020Different2,2013Quantum, 10.1140/epjp/s13360-021-02009-3,2010Noise,lizhao2021}.

Previous work on the EET in pigment chains with an arbitrary number of sites\cite{2009Optimization,2012Coherence,2012Dephasing,Elisabet2017} assumed that the pigments were arranged in a similar pattern in the molecular chain and respond to a uniform environment in the pigment-protein complexes\cite{2010chinNoise,PhysRevE.103.042124,2021Simulating}. Following that, numerous theoretical models for studying the EET were established.  Some authors discovered that spatially correlated fluctuations improved charge delocalization and charge transfer rates in a system of units with uniform site-energies, and that temporal correlations of site-energies were important in determining the coherent-incoherent transition in a coarse-grained theory of biological charge transfer\cite{2016Coarse}. The correlated fluctuations between site-energy and inter-site electronic couplings boosted or suppressed coherence and transfer rate in the two-state model for reaction center system\cite{2012Influencehuo} with accessory bacteriochlorophyll and bacteriopheophytin. A differential entropy dependent diffusion and charge density equation\cite{2019EffectNavamani} were suggested to elucidate the link between the dynamics of site-energy disorder and charge transport. Furthermore, due to site energy variations, dispersion transport was exhibited together with the extended $\pi$-stacked molecules, as well as the current density of distinct site-energy differences highlighted the validity and limitations of the Einstein relation\cite{2019EffectNavamani}. The effects of spatial correlations on EET dynamics were recently examined in a simulated donor-acceptor model, and it demonstrated that spatial correlations of bath fluctuations play a role in safeguarding coherence in the decay of coherence dynamics\cite{2021Roledu}.

However, few studies have focused on the effect of quantum fluctuations at different site-energies on EET as a result of the interaction between site-energy and the ambient environment. The assumption that all pigments in pigment chains are equal, we hypothesize, may lead to the omission of some quantum aspects of EET in natural photosynthetic systems. As a result, linked quantum fluctuations of site-energies were introduced among the densely packed chromophores in the protein or solvent environment in this study. To gain a better understanding of the EET process, we investigate EET in two neighboring pigment molecules and analyze EET features and transport efficiency using j-V characteristics and power in real photosynthetic systems simulated by a quantum heat engine (QHE).

With the goal in mind, this paper is organized as follows: In Sec. II, we introduce the corrected fluctuations at different site-energies in two adjacent pigments via a QHE frame, and theoretically describe EET process. Then, we discuss and analyze the influence on the EET in Sec.III. We give some remarks and point out the significance of the current work in Sec. IV. And then we conclude the work in Sec. V.

\section{Two adjacent pigments with correlated fluctuations in site energy }
\subsection{Theoretical model for two adjacent pigments}

\begin{figure}[htp]
\center
\subfigure[]{\includegraphics[width=0.45\columnwidth]{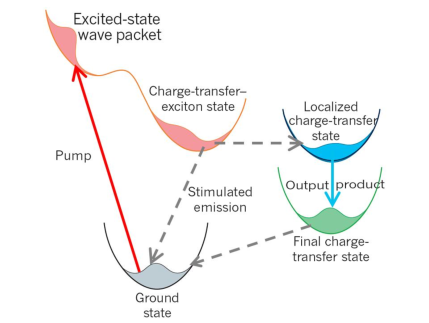}}
\hspace{0.10in}
\subfigure[]{\includegraphics[width=0.45\columnwidth]{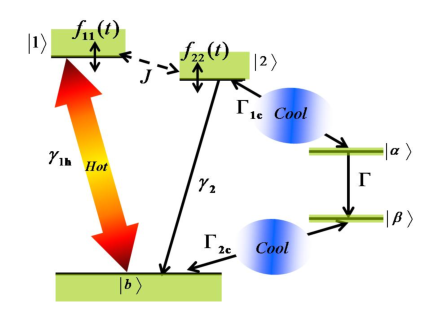}}
\caption{(Color online) (a) The energy level scheme depicted by wave packet. Excited-state wave packet can be produced by solar photon absorption, and the subsequent dynamics of charge-transfer exciton state and output charge-separated product are probed by intersite coupling and stimulated emission processes. (b) The intuitive energy-level diagram of two pigment molecules corresponding to (a). }
\label{Fig.1}
\end{figure}

In contrast to previous work, two adjacent pigment molecules will be evaluated in this work to clarify the different roles of sites in the EET process.
Wave packet motion encompasses both coherent and non-coherent transfer motion in this proposed model. As shown in Fig.\ref{Fig.1}(a), the pump can generate the initially excited state wave packet due to an external source via photon absorption. And, as the exciton passes through two pigments via intersite coherent coupling, it couples to phonons produced by vibrational motion of sites to form a charge-transfer exciton state (see in Fig.\ref{Fig.1}(a)). This model was used to describe the EET in photosynthetic complexes\cite{1997Exciton,2006How,RevModPhys.90.035003}, organic semiconductor\cite{2007Charge2}, and quantum networks\cite{PhysRevLett.122.050501}. Furthermore, non-coherent transfer from the charge-transfer exciton state to the localized charge-transfer state and the ground state, as well as from the final charge-transfer state to the ground state, is depicted by gray dashed arrows (as shown in Fig.\ref{Fig.1}(a)). The transition between the localized charge-transfer state and the final charge-transfer state produces the external output quantum yield. In the following analysis, the photon-absorbing pigment molecule is referred to as Site 1, and the adjacent pigment molecule in the charge-transfer excited state is referred to as Site 2(as shown in Fig.\ref{Fig.1} (a)).

Corresponding to Fig.\ref{Fig.1}(a), Fig.\ref{Fig.1}(b) is the energy level diagram described by quantum heat engine (QHE), in which the EET process is depicted with fluctuating in site energies $f_{ii}(t)$, $i$=1,2 as well as with a fluctuating transfer matrix element $f_{12}(t)$.For the sake of accurate solution, the fluctuations are treated as classical Gaussian Markov processes, i.e., a delta function time dependence is chosen for the correlation functions of the fluctuations in this proposed pigment dimer\cite{PhysRev.129.597,1972The,H1973An,2010Effect2}. As a result, fluctuations in site energies can be described by the correlation function $\langle f_{ii}(t)f_{ii}(t_{0})\rangle$=$\gamma_{ii;ii}\delta(t-t_{0})$,(i=1,2), and fluctuations in $f_{12}(t)$ by the correlation function $\langle f_{12}(t)f_{12}(t_{0})\rangle$=$\gamma_{12;12}\delta(t-t_{0})$. As a result, the correlated fluctuations between different sites can be represented as $ \langle f_{11}(t)f_{22}(t_{0})\rangle$=$\gamma_{11;22}\delta(t-t_{0})$. Furthermore, the correlations between fluctuations in a site energy and fluctuations in the energy transfer matrix element can be considered as $\langle f_{ii}(t)f_{12}(t_{0})\rangle$=$\gamma_{ii;12}\delta(t-t_{0}),(i=1,2)$. Thus, the correlated fluctuations' influence on EET can be explicitly clarified by the aforementioned fluctuation parameters.

The double arrows with warm tone depict the process of photon absorption at rate $\gamma_{1h}$, during which the fluctuation is valued by $\gamma_{11;11}$ due to the vibrational motion of site $|1\rangle$  coupling to the ambient phonons, and the transitions $|b\rangle$ $\leftrightarrow$ $|1\rangle$ denote the harvesting-light processes, as shown in Fig.\ref{Fig.1}(b). In this proposed QHE model, the hot bath is the sun environment. As a result of the inter-site coupling strength $J$, the excitations are transported through two pigments, and the fluctuation $\gamma_{22;22}$ forms in the charge-transfer exciton state $|2\rangle$ on the second pigment molecule. Transfers from the charge-transfer exciton state $|2\rangle$ to the localized charge-transfer state  $|\alpha\rangle$ and the ground state $|b\rangle$ can occur via stimulated emission at rates of $\Gamma_{1c}$  and  $\gamma_{2}$, respectively
And the ground state's site energy $|b\rangle$ is set to zero. The localized charge-transfer state $|\alpha\rangle$ is transmitted in the form of output photoelectric energy to the final charge-transfer state $|\beta\rangle$ and then back to the ground state $|b\rangle$  at rates $\Gamma$ and $\Gamma_{2c}$, respectively. In this QHE model, the transitions $|2\rangle$ $\leftrightarrow$ $|\beta\rangle$ and $|\beta\rangle$ $\leftrightarrow$ $|b\rangle$  interact with the ambient environment to form the Cool bath. The external terminal can realize the photosynthesis output via the transition $|\alpha\rangle$ $\leftrightarrow$ $|\beta\rangle$. The Hamiltonian in the proposed model's site basis is written as,

\begin{eqnarray}
\hat{H}=\hat{H}_{S}+\hat{H}_{B}+\hat{H}_{int},   \label{1}
\end{eqnarray}

\noindent in which $\hat{H}_{S}$ represents the Hamiltonian of the five-level QHE system and it can be written as follows,

\begin{eqnarray}
 &H_{S}=&\sum_{i=1,2}\varepsilon_{i}|i\rangle\langle i|+J (|1\rangle\langle 2|+|2\rangle\langle 1|)+\sum_{i=1,2}\sum_{j=1,2}f_{ij}(t)|i\rangle\langle j|\nonumber\\&& + \sum_{n=\alpha,\beta,b}\varepsilon_{n}|n\rangle\langle n|,      \label{2}
\end{eqnarray}

\noindent where $\varepsilon_{i}$ denotes the energy of the ith site and $J$ denotes the coupling coefficient between two adjacent site states $|1\rangle$ and $|2\rangle$. The Hamiltonian of the ambient bath interplaying with the proposed system is described by $\hat{H}_{B}$ in Eq(\ref{1}), and its expression is described by,

\begin{eqnarray}
\hat{H}_{B}=\sum_{k}\hbar\omega_{k}\hat{a}^{\dag}_{k}\hat{a}_{k}+\sum_{q}\hbar\omega_{q}\hat{b}^{\dag}_{q}\hat{b}_{q},      \label{3}
\end{eqnarray}

\noindent with $\hat{a}^{\dag}_{k}$($\hat{a}_{k}$) being the photons' creation (annihilation) operator and the k-th noninteracting photon mode frequency $\omega_{k}$. $\hat{b}^{\dag}_{q}$($\hat{b}_{q}$) is the ambient environment phonon creation (annihilation) operator with its q-th harmonic oscillator mode frequency.

The third part of Eq(\ref{1}) describes the Hamiltonian $\hat{H}_{int}$ interaction between the system and the surrounding environment. According to the rotating-wave approximations \cite{lizhao2021,ZHAO2020106329,10.1140/epjp/s13360-021-02009-3}, $\hat{H}_{int}$  is as follows,

\begin{align}
\hat{H}_{int}=\hat{V}_{h}+\hat{V}_{c},    \label{4}
\end{align}

\noindent with the following elements,

\begin{eqnarray}
&\hat{V}_{h}=&\sum_{i=1,2}\sum_{k}\hbar(g_{i,k}^{(h)}\hat{\sigma}_{b,i}\otimes\hat a_{i,k}^{\dag}+g_{i,k}^{(h)\ast}\hat{\sigma}_{b,i}^{\dag}\otimes\hat a_{i,k}), \\    \label{5}
&\hat{V}_{\Gamma_{c}}=&\sum_{q}[\hbar(g_{\alpha,2}^{(c)}\hat{\sigma}_{\alpha,2}\otimes\hat b_{c,q}^{\dag}+g_{\alpha,2}^{(c)\ast}\hat{\sigma}_{\alpha,2}^{\dag}\otimes\hat b_{c,q})\nonumber\\&&+\hbar(g_{b,\beta}^{(c)}\hat{\sigma}_{b,\beta}\otimes\hat b_{c,q}^{\dag}+g_{b,\beta}^{(c)\ast}\hat{\sigma}_{b,\beta}^{\dag}\otimes\hat b_{c,q})],  \label{6}
\end{eqnarray}

\noindent where $g_{i,k}^{(h)}$ denotes the coupling strength of the i-th pigment to the k-th Hot bath mode, $g_{\alpha,2}^{(c)}$ ($g_{b,\beta}^{(c)}$) describes the coupling strength between the transition $|2\rangle$ $\leftrightarrow$ $|\alpha\rangle$ ($|\beta\rangle$ $\leftrightarrow$ $|b\rangle$) and the q-th Cold bath mode. And $\hat{a}_{i,k}^{\dag}$, $\hat{b}_{c,q}^{\dag}$($\hat{a}_{i,k}$, $\hat{b}_{c,q}$) are the creation (annihilation) operations of the Hot and Cold bath, respectively. The lowering operators are defined as: $\hat{\sigma}_{b,i}=|b\rangle\langle i|$, $\hat{\sigma}_{b,\beta}=|b\rangle\langle \beta|$, $\hat{\sigma}_{\alpha,i}=|\alpha\rangle\langle i|$$_{(i=1,2)}$.

To evaluate the different roles of correlated fluctuations in site-energy in EET, we will focus on fluctuations in site-energies characterized by the correlation function $\gamma_{ii; ii}\delta(t-t_{0})$,(i=1,2), rather than the correlated fluctuating transfer matrix elements and the correlated fluctuations between different sites, which are related to the coherence or incoherence behaviors in the exciton transfer molecular chain and the site-energy and the site-energy fluctuation correlations on the evolution of quantum coherence have been discussed in previous work, see Ref. \cite{PhysRev.129.597,1972The,H1973An,2010Effect2,2012Influencehuo,PhysRevLett.122.050501,2019EffectNavamani}. In this paper, we will focus on the photosynthesis properties aided by quantum fluctuations in various site energies, and we will clarify the various roles of quantum fluctuations in EET corresponding to their site energies. A more complete understanding of the influence of quantum fluctuations in various site-energies on EET will be achieved.

\subsection{Derivation of the dynamics of the reduced density matrix}

In the chronological time-ordering prescription\cite{2010Effect2,PhysRevA.17.1988}, the general second-order treatment of the dynamics of the reduced density matrix in the weak coupling regime can be re-written in the Schr$\ddot{o}$dinger picture,

\begin{widetext}
\begin{eqnarray}
&\frac{d\hat{\rho}}{dt}=&-i[\hat{H}_{S},\hat{\rho}]-\int^{t}_{0}d\tau C(\tau)\hat{\rho}(t-\tau)+L_{H}\hat{\rho}+L_{2}\hat{\rho}+L_{\Gamma}\hat{\rho}+L_{\Gamma_{1c}}\hat{\rho}+L_{\Gamma_{2c}}\hat{\rho},    \label{7}
\end{eqnarray}
\end{widetext}
\noindent where $\hat{\rho}$ is the reduced density matrix of the system, and with $C(\tau)$ being defined as,
\begin{eqnarray}
&C(\tau)=&\langle\mathcal{L}_{int}exp[-i\mathcal{L}_{0}\tau]\mathcal{L}_{int}exp[i\mathcal{L}_{0}\tau]\rangle_{B},    \label{8}
\end{eqnarray}
\noindent with $\mathcal{L}_{int}$=$[\hat{H}_{int},\cdot]$, $\mathcal{L}_{0}$=$[\hat{H}_{S}+\hat{H}_{B},\cdot]$. Therefore, the explicit expression of $C(\tau)\rho(t-\tau)$ in Eq.(\ref{7}) can be written in the Schr$\ddot{o}$dinger picture as follows,

\begin{widetext}
\begin{eqnarray}
&C(\tau)\rho(t-\tau)\!=\!&Tr\langle[ \hat{H}_{int},\hat{U}_{0}(\tau)[\hat{H}_{int},\hat{\rho}(t-\tau)]\hat{U}^{\dag}_{0}(\tau)]\rangle_{B},    \label{9}
\end{eqnarray}
\end{widetext}

\noindent with the unitary operator $\hat{U}_{0}(\tau)$=$exp[-i(\hat{H}_{S}+\hat{H}_{B})t]$. And the other five Lindblad-type superoperator terms in Eq.(\ref{7}) are deduced with the following expressions below:

\begin{widetext}
\begin{eqnarray}
&L_{H}\hat{\rho}\!=\!&\frac{\gamma_{1h}}{2}[(n_{1h}+1)(\hat{2\sigma}_{b,1}\hat{\rho}\hat{\sigma}_{b,1}^{\dag}-\hat{\sigma}_{b,1}^{\dag}\hat{\sigma}_{b,1} \hat{\rho}-\hat{\rho}\hat{\sigma}_{b,1}^{\dag}\hat{\sigma}_{b,1}) +n_{1h}(2\hat{\sigma}_{b,1}^{\dag}\hat{\rho}\hat{\sigma}_{b,1} -\hat{\sigma}_{b,1}\hat{\sigma}_{b,1}^{\dag}\hat{\rho}-\hat{\rho}\hat{\sigma}_{b,1}\hat{\sigma}_{b,1}^{\dag})], \nonumber\\
&L_{2}\hat{\rho}\!=\!&\frac{\gamma_{2}}{2}[(n_{2}+1)(\hat{2\sigma}_{b,2}\hat{\rho}\hat{\sigma}_{b,2}^{\dag}-\hat{\sigma}_{b,2}^{\dag}\hat{\sigma}_{b,2}\hat{\rho}-\hat{\rho}\hat{\sigma}_{b,2}^{\dag}\hat{\sigma}_{b,2}],\nonumber\\
&L_{\Gamma}\hat{\rho}\!=\!&\frac{\Gamma}{2}[2\hat{\sigma}_{\beta\alpha}\hat{\rho}\hat{\sigma}_{\beta\alpha}^{\dag}-\hat{\rho}\hat{\sigma}_{\beta\alpha}^{\dag}\hat{\sigma}_{\beta\alpha}
                                                                           -\hat{\sigma}_{\beta\alpha}^{\dag}\hat{\sigma}_{\beta\alpha}\hat{\rho}],                             \label{10}\\
&L_{\Gamma_{1c}}\hat{\rho}\!=\!&\frac{\Gamma_{1c}}{2}[(n_{1c}+1)(2\hat{\sigma}_{\alpha,2}\hat{\rho}\hat{\sigma}_{\alpha,2}^{\dag}-\hat{\sigma}_{\alpha,2}^{\dag}\hat{\sigma}_{\alpha,2}
\hat{\rho}-\hat{\rho}\hat{\sigma}_{\alpha,2}^{\dag}\hat{\sigma}_{\alpha,2})+n_{1c}(2\hat{\sigma}_{\alpha,2}^{\dag}\hat{\rho}\hat{\sigma}_{\alpha,2}-\hat{\sigma}_{\alpha,2}\hat{\sigma}_{\alpha,2}^{\dag}\hat{\rho}-\hat{\rho}\hat{\sigma}_{\alpha,2}\hat{\sigma}_{\alpha,2}^{\dag})],\nonumber\\
&L_{\Gamma_{2c}}\hat{\rho}\!=\!&\frac{\Gamma_{2c}}{2} [(n_{2c}+1)(2\hat{\sigma}_{b,\beta}\hat{\rho}\hat{\sigma}_{b,\beta}^{\dag}-\hat{\rho}\hat{\sigma}_{b,\beta}^{\dag}\hat{\sigma}_{b,\beta}-\hat{\sigma}_{b,\beta}^{\dag}\hat{\sigma}_{b,\beta}\hat{\rho})
+n_{2c}(2\hat{\sigma}_{\beta,b}\hat{\rho}\hat{\sigma}_{\beta,b}^{\dag}-\hat{\rho}\hat{\sigma}_{\beta,b}^{\dag}\hat{\sigma}_{\beta,b}-\hat{\sigma}_{\beta,b}^{\dag}\hat{\sigma}_{\beta,b}\hat{\rho})].\nonumber
\end{eqnarray}
\end{widetext}

\noindent where $\gamma_{1h}$($\gamma_{2}$) denotes stimulated emission rate from $|1\rangle$ to $|b\rangle$ ($|2\rangle$ to $|b\rangle$) with $n_{1h}$=$[exp(\frac{\varepsilon_{1b}}{K_{B}T_{s}})-1]^{-1}$ ($n_{2}$=$[exp(\frac{\varepsilon_{2b}}{K_{B}T_{s}})-1]^{-1}$). And $\varepsilon_{1b}$ ($\varepsilon_{2b}$) is the corresponding energy difference, $T_{s}$ denotes the sun temperature. $\Gamma_{1c}$ ($\Gamma_{2c}$) represents the transition rate from $|2\rangle$$\leftrightarrow$ $|\alpha\rangle$ ($|\beta\rangle$ $\leftrightarrow$ $|b\rangle$) with population: ${n}_{1c}$=$[exp\frac{\varepsilon_{2\alpha}}{K_{B}T_{a}}-1]^{-1}$ ($n_{2c}$=$[exp\frac{\varepsilon_{b\beta}}{K_{B}T_{a}}-1]^{-1}$), where $T_{a}$ represents the ambient temperature and $\varepsilon_{2\alpha}$ ($\varepsilon_{b\beta}$) is the energy difference. $L_{\Gamma}\hat{\rho}$ describes a process that the system decays from state $|\alpha\rangle$ to state $|\beta\rangle$, which leads to the current proportional to the relaxation rate $\Gamma$ as defined before.

As previously stated, we simply define $C_{ij,i'j'}(\tau)$=$C^{*}_{ij,i'j'}(\tau)$=$\gamma_{ij,i'j'}(\tau)$ for the classic white noise assumption. To put it another way, we replace $f_{ij}(t)$ written in terms of phonon operators with time dependent terms with zero average, which is a Gaussian Markov process with a delta function correlation in time and defines the covariance matrix of the underlining Gaussian process. We assume that all of the aforementioned fluctuations are Gaussian stochastic processes with second-order correlation functions proportional to $\delta(t-t_{0})$. As a result, the Eq.(\ref{7}) can be deduced from the most general master equation,

\begin{widetext}
\begin{eqnarray}
&\frac{d\hat{\rho}_{ij}(t)}{dt}=&-i[\hat{H}_{S},\hat{\rho}_{ij}(t)]-\sum_{i',j'}[\gamma_{ii',j'j}\hat{\rho}_{i'j'}(t)+\gamma_{j'j,ii'}\hat{\rho}_{i'j'}(t)-\gamma_{ij',i'j'}\hat{\rho}_{i'j}(t)-\gamma_{i'j,j'i'}\hat{\rho}_{ij'}(t)]
\nonumber\\&&+L_{H}\hat{\rho}+L_{2}\hat{\rho}+L_{\Gamma}\hat{\rho}+L_{\Gamma_{1c}}\hat{\rho}+L_{\Gamma_{2c}}\hat{\rho},    \label{11}
\end{eqnarray}
\end{widetext}

Invoking the Weisskopf-Wigner approximation, the reduced matrix elements corresponding to Eq.(\ref{11}) can be written as follows,

\begin{widetext}\begin{eqnarray}
&\dot{\rho}_{11}=&-\gamma_{1h}[(n_{1h}+1)\rho_{11}-n_{1h}\rho_{bb}]+iJ(\rho_{12}-\rho_{21})+2\gamma_{12;12}(\rho_{22}-\rho_{11})+(\gamma_{11;12}-\gamma_{22;12})(\rho_{12}+\rho_{21}),\nonumber\\
&\dot{\rho}_{22}=&-\gamma_{2}[(n_{2}+1)\rho_{22}]-\gamma_{2c}[(n_{1c}+1)\rho_{22}-n_{1c}\rho_{\alpha\alpha}]-iJ(\rho_{12}-\rho_{21})-2\gamma_{21;21}(\rho_{22}-\rho_{11})
\nonumber\\&&-(\gamma_{11;12}-\gamma_{22;12})(\rho_{12}+\rho_{21}),\nonumber\\
&\dot{\rho}_{12}=& i\rho_{12}\varepsilon_{12}-iJ(\rho_{22}-\rho_{11})-\frac{\gamma_{1h}}{2}(n_{1h}+1)\rho_{12}-\frac{\gamma_{2}}{2}(n_{2}+1)\rho_{12}-\frac{\gamma_{2}}{2}(n_{1c}+1)\rho_{12}
 \\ &&-(\gamma_{11;11}+\gamma_{22;22}-2\gamma_{11;22})\rho_{12}-2\gamma_{12;12}(\rho_{12}-\rho_{21})-(\gamma_{22;12}-\gamma_{11;12})(\rho_{11}-\rho_{22}),\nonumber\\
&\dot{\rho}_{\alpha\alpha}=&\Gamma_{1c}[(n_{2c}+1)\rho_{22}-n_{1c}\rho_{\alpha\alpha}]-\Gamma\rho_{\alpha\alpha},\nonumber\\
&\dot{\rho}_{\beta\beta}=&\Gamma\rho_{\alpha\alpha}-\Gamma_{2c}(n_{2c}+1)\rho_{\beta\beta}+\Gamma_{2c}n_{2c}\rho_{bb} \nonumber
\end{eqnarray}\end{widetext}

\noindent where $\rho_{ii}$ represents the diagonal element and $\rho_{ij}$ describes the non-diagonal element, and in the following discussion we set $\gamma_{12;12}$=$\gamma_{21;21}$. Consider the Boltzmann distribution on the upper and lower levels $|\alpha \rangle$ and  $|\beta \rangle$ with the populations $\rho_{\alpha\alpha}$=$exp(-\frac{E_{\alpha}-\mu_{\alpha}}{k_{B}T_{a}})$ and $\rho_{\beta\beta}$=$exp(-\frac{E_{\beta}-\mu_{\beta}}{k_{B}T_{a}})$. An effective voltage $V$ can be introduced as a load drop to the transition $|\alpha\rangle\longrightarrow|\beta\rangle$, utilizing the difference in chemical potentials $\mu_{\alpha}$ and $\mu_{\beta}$ via the expression, $eV =\varepsilon_{\alpha\beta}+K_{B}T_{a}\ln\frac{\rho_{\alpha\alpha}}{\rho_{\beta\beta}}$. The elementary electric charge is denoted by $e$. As a result, the EET can be quantified using the current $ j=e\Gamma\rho_{\alpha\alpha}$ definition. And the proposed five-level QHE system's output power can be calculated using $P=jV$.

\begin{table}
\begin{center}
\caption{This model's parameters are listed below.}
\label{Table.1}
\vskip 0.2cm\setlength{\tabcolsep}{0.5cm}
\begin{tabular}{ccc}
\hline
\hline
                                               & Values                  & Units \\
\hline
\(\gamma_{1h}\)                                & $8.5\times 10^{-6}$     & Hz    \\
\(\gamma_{2}\)                                 & $4.5\times 10^{-6}$     & Hz    \\
\(\gamma_{12;12}\)                             & $1.5\times 10^{-6}$     & Hz    \\
\(\varepsilon_{2\alpha}\)                      & 0.2                     & eV    \\
\(\varepsilon_{ b\beta}\)                      & 0.2                     & eV    \\
\(T_{a}\)                                      & 0.026                   & eV    \\
\(n_{1h}\)                                     & 6000                    &       \\
\(n_{2}\)                                      & 5000                    &       \\
\hline
\hline
\end{tabular}
\end{center}
\end{table}

\section{Results and analysis}

We discuss the quantitative relationship between EET and correlated fluctuations at different sites using Current-voltage ($j$-$V$) characteristics curves and output power curves to highlight the influence of correlated fluctuations at different site-energies on EET. The correlated fluctuating transfer matrix elements and the correlated fluctuations between different site-energies are not considered in the following discussion because they involve quantum coherence and have been discussed in detail in some previous work\cite{PhysRev.129.597,1972The,H1973An,2010Effect2,2012Influencehuo,PhysRevLett.122.050501,ZHAO2020106329,2019EffectNavamani}. As a result, some parameters were set in this proposed model for the correlated fluctuating transfer matrix elements and the correlated fluctuations between different site-energies, while we focused on the regulatory features of fluctuations at different sites to reveal their distinct influences on EET while interacting with the ambient environment. The Table \ref{Table.1} lists some of the common parameters used in this model.

\begin{figure}[htp]
\center
\includegraphics[width=0.45\columnwidth]{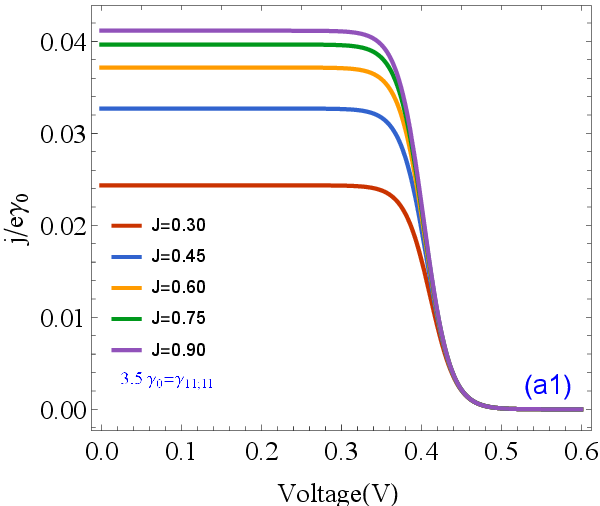}\includegraphics[width=0.45\columnwidth]{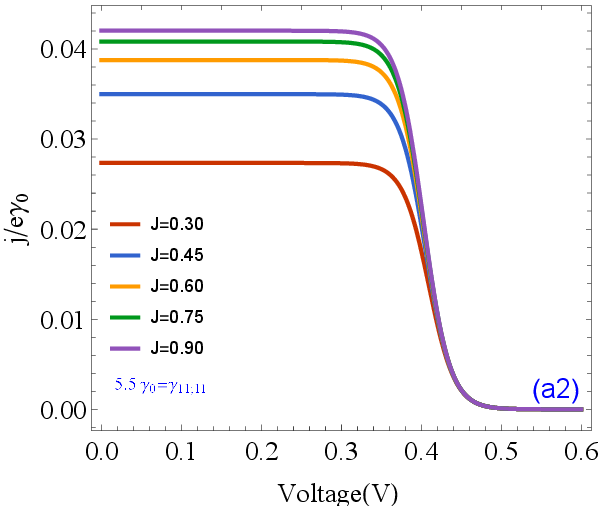}
\hspace{0in}%
\includegraphics[width=0.45\columnwidth]{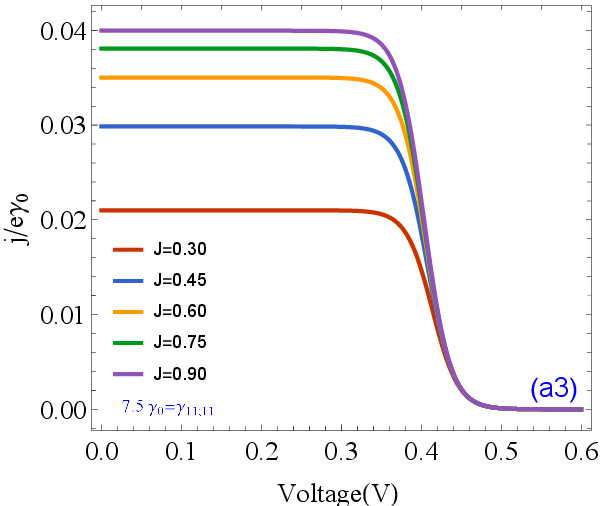 }\includegraphics[width=0.45\columnwidth]{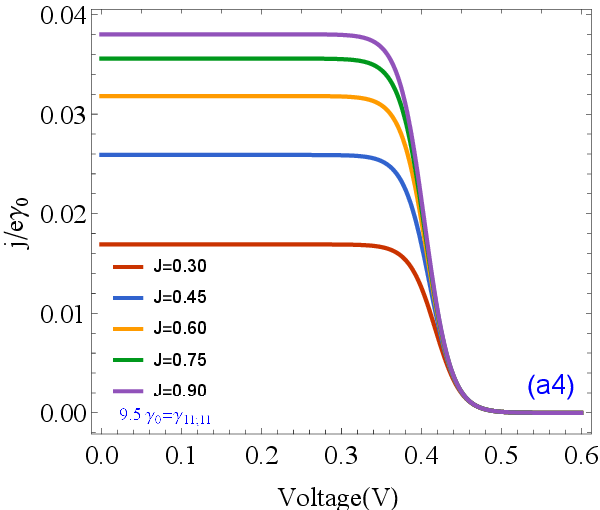 }
\caption{(Color online) Current-voltage ($j$-$V$) characteristics assisted by the correlation fluctuation ($\gamma_{11;11}$) of Site 1 with different coupling coefficients $J$. Other parameters are $\gamma_{11;12}$ - $\gamma_{22;12}$=$2.2\gamma_{0}$, $\gamma_{22;22}$=$2.8\gamma_{0}$, $\gamma_{11;22}$=$3.2\gamma_{0}$, $\Gamma_{1c}$=$0.48\gamma_{0}$, $\Gamma_{2c}$=$0.18\gamma_{0}$, $\Gamma$=$0.24\gamma_{0}$, $\varepsilon_{\alpha\beta}$=$0.6 eV$, $\gamma_{0}$=0.8$Hz$. }
\label{Fig.2}
\end{figure}

\begin{figure}[htp]
\center
\includegraphics[width=0.45\columnwidth]{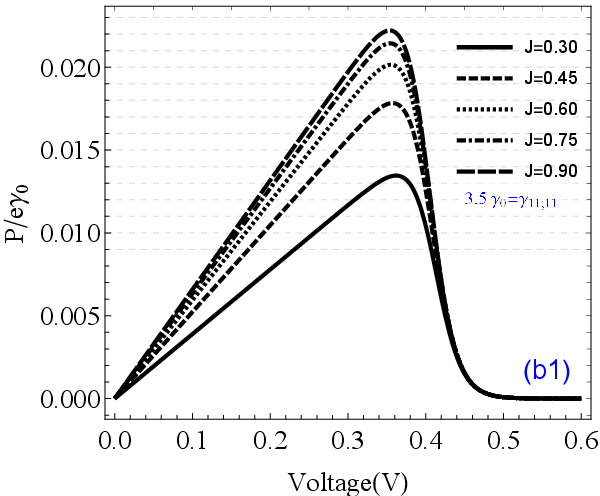}\includegraphics[width=0.45\columnwidth]{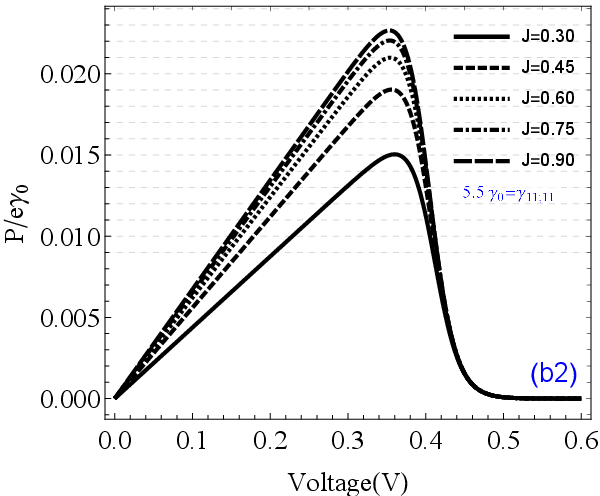}
\hspace{0in}%
\includegraphics[width=0.45\columnwidth]{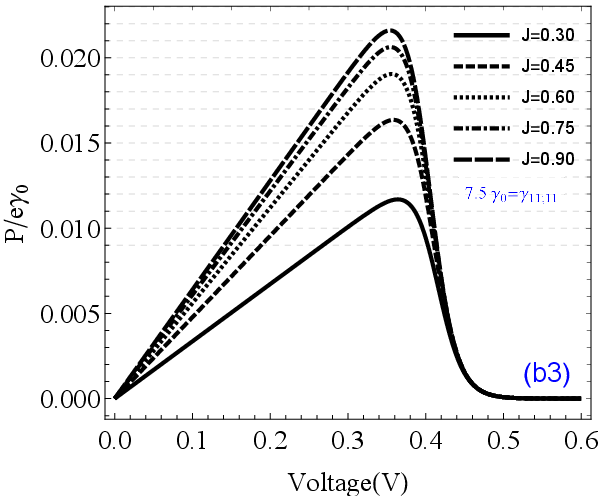}\includegraphics[width=0.45\columnwidth]{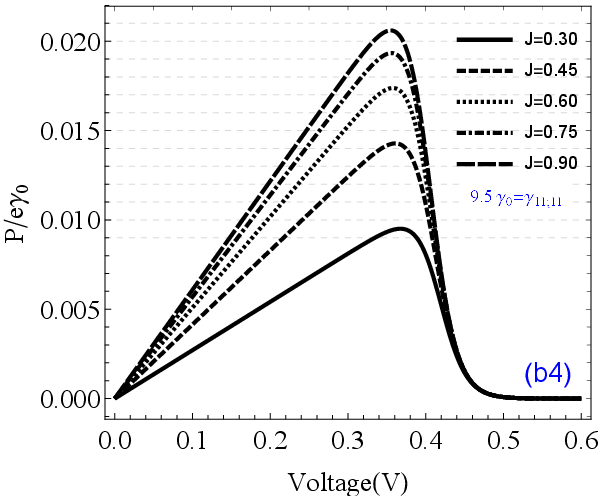}
\caption{(Color online) The output power $P$ versus voltage $V$ assisted by the correlation fluctuation ($\gamma_{11;11}$) of Site 1 with different coupling coefficients $J$. All the other parameters are the same to those in Fig.\ref{Fig.2}.}
\label{Fig.3}
\end{figure}

Ref.\cite{2010Effect2} investigated the dynamics of exciton populations and coherence as affected by correlations of site-energies fluctuations and provided insight into how these correlations affect the evolution of exciton populations and coherence among the closely packed chromophores in the protein environment. The effects of fluctuations at different sites on EET, however, were not differentiated. EET regulated by fluctuations at Site 1 is plotted in Fig.\ref{Fig.2} and Fig.\ref{Fig.3} with varying coupling intensities $J$. In Fig.\ref{Fig.2} and Fig.\ref{Fig.3}, the correlated fluctuations at Site 1 are valued by $\gamma_{11;11}$for simplicity. The short-circuit currents (see from the horizontal lines) increase with the increments of inter-site coherent coupling intensities $J$, as shown by the curves in Fig.\ref{Fig.2}. At the same time, we can see that the short-circuit current ascends with increasing fluctuation intensity from $\gamma_{11;11}$=3.5$\gamma_{0}$ in (a1) to $\gamma_{11;11}$=5.5$\gamma_{0}$ in (a2), and then descends with $\gamma_{11;11}$=7.5$\gamma_{0}$ in (a3) to $\gamma_{11;11}$=9.5$\gamma_{0}$(a4). It is noted that the EET begins to wane as the correlated fluctuation intensity $\gamma_{11;11}$=7.5$\gamma_{0}$ at Site 1. When $J$ =0.30 (corresponding to the red curves), the short-circuit current first increases from about 0.024 to 0.028 (as shown in (a1) and (a2)), and then decreases as the fluctuation intensity increases, from about 0.022 to 0.018 (as shown in (a3) and (a4).

The inter-site coupling intensity $J$ in packed chromophores can be used to describe the density of packed chromophores. The increasing inter-site coupling among the closely packed chromophores can enhance wave packet coherent motion, resulting in an increase in EET. Similarly, some different photoelectric conversion theoretical models\cite{lizhao2021,ZHONG2021104094}, have demonstrated similar physical behaviors. The quantum coherence has been shown to improve photosynthetic properties \cite{2010Quantum1,2010Coherently,doi:10.1021/acs.jpcb.0c10719}. As a result, we hypothesize that the quantum coherence generated by interactions between two adjacent sites can be enhanced by the appropriate quantum correlated fluctuation, whereas too strong correlated fluctuation destroys the quantum coherence. As a result, the underlying physical regime of EET in this proposed model is quantum coherence, which produces the first ascending and then descending EET.

The output power confirms the regulated nature of the corrected fluctuation intensities $\gamma_{11;11}$ on EET in Fig.\ref{Fig.3}. The peak output powers ascend with 0.15 increments in the coupling intensity $J$, as shown in Fig.\ref{Fig.3} (b1). It also mentions that the peak powers follow the same pattern as in Fig.\ref{Fig.2}, climbing up and then declining due to $\gamma_{11;11}$ increasing from $\gamma_{11;11}$=3.5$\gamma_{0}$ in (b1), $\gamma_{11;11}$=5.5$\gamma_{0}$ in (b2), $\gamma_{11;11}$=7.5$\gamma_{0}$ in (b3), and $\gamma_{11;11}$=9.5$\gamma_{0}$ in (b4) with  $J$=0.90.

\begin{figure}[htp]
\center
\includegraphics[width=0.45\columnwidth]{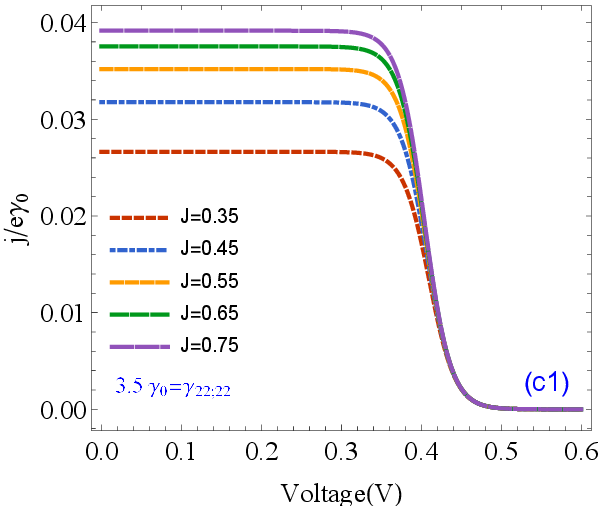}\includegraphics[width=0.45\columnwidth]{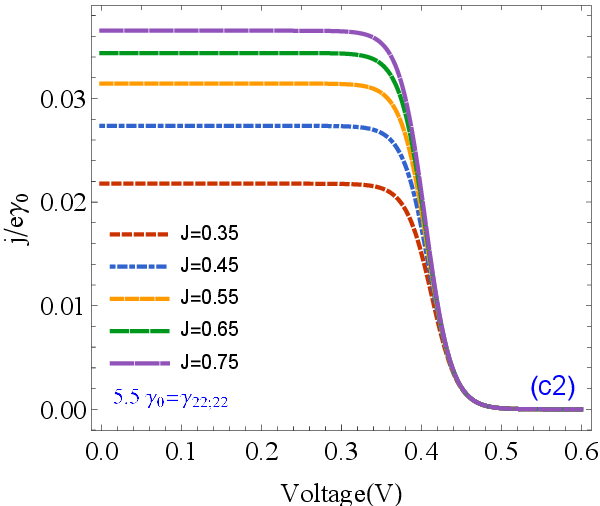}
\hspace{0in}%
\includegraphics[width=0.45\columnwidth]{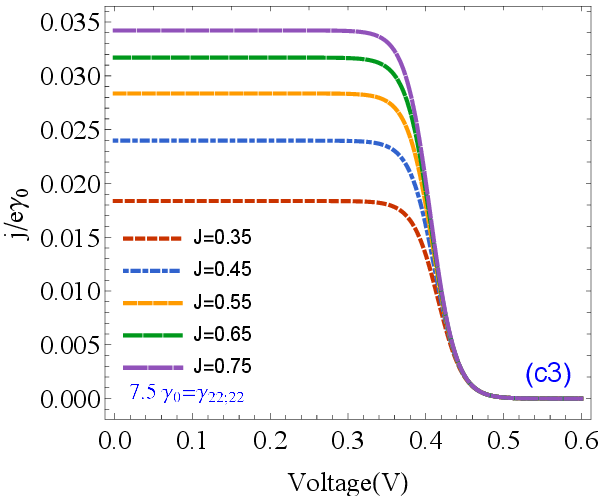}\includegraphics[width=0.45\columnwidth]{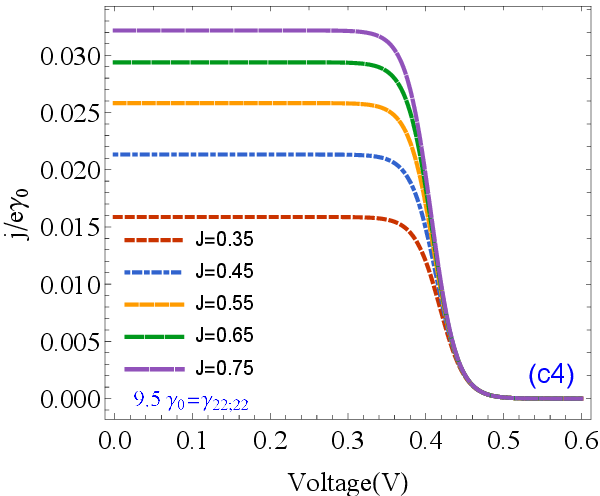}
\caption{(Color online) Current-voltage ($j$-$V$) characteristics assisted by the correlation fluctuation $\gamma_{22;22}$ of Site 2 with different coupling coefficients $J$. $\gamma_{11;11}$=$6\gamma_{0}$, and all the other parameters are the same to those in Fig.{\ref{Fig.2}}.}
\label{Fig.4}
\end{figure}

In previous research\cite{2010chinNoise,PhysRevE.103.042124,2021Simulating}, the pigments were assumed to be arranged in a similar pattern in the molecular chain and responding to a uniform environment in pigment-protein complexes. However, we hypothesize that different pigment molecules may play different roles in EET due to their different responses to interactions with the surrounding environment. As a result, we believe that fluctuation at different sites may play different roles in the interaction between photosynthetic light-harvesting complexes and the surrounding environment, and the influence of corrected fluctuation at Site 2 attracts our interest.

\begin{figure}[htp]
\center
\includegraphics[width=0.45\columnwidth]{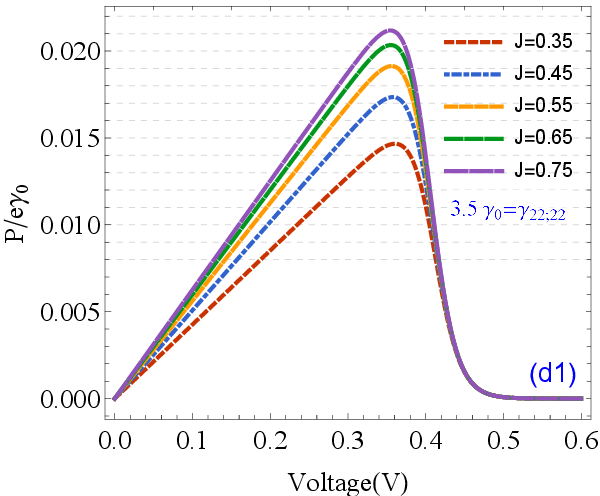}\includegraphics[width=0.45\columnwidth]{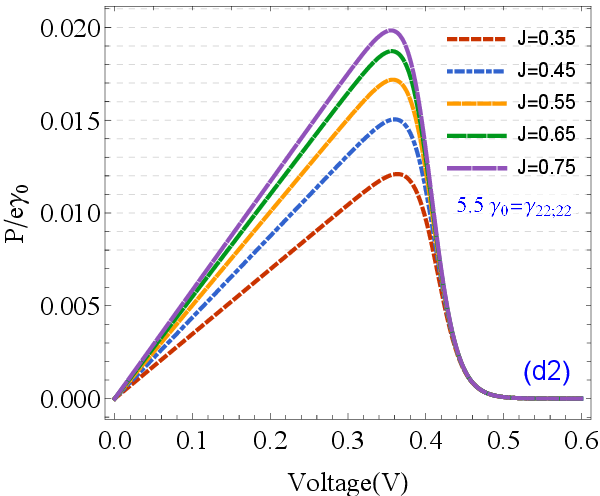}
\hspace{0in}%
\includegraphics[width=0.45\columnwidth]{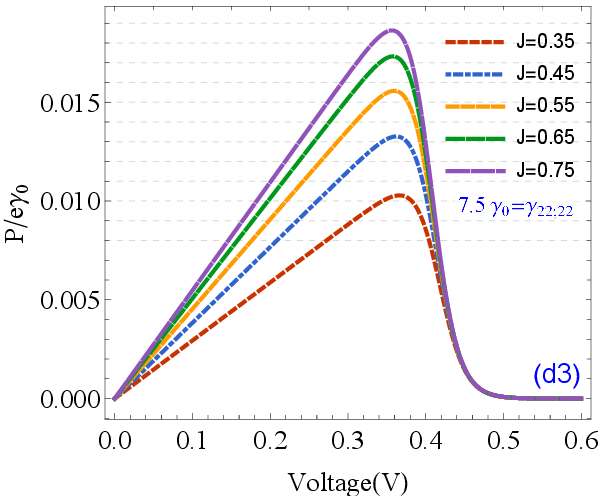}\includegraphics[width=0.45\columnwidth]{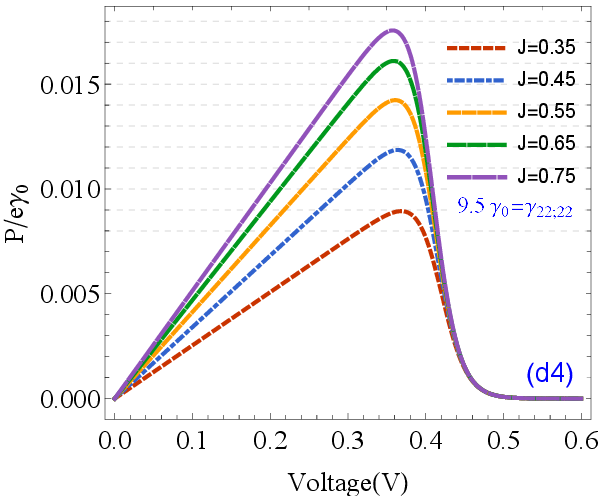}
\caption{(Color online) The output power $P$ versus voltage $V$ assisted by the correlation fluctuation $\gamma_{22;22}$ of Site 2 with different coupling coefficients $J$. All the other parameters are the same to those in Fig.\ref{Fig.4}.}
\label{Fig.5}
\end{figure}

Similarly, the current-voltage ($j$-$V$) characteristics and output power P of Site 2 are plotted in Fig.\ref{Fig.4} and Fig.\ref{Fig.5}, which is used to access EET features with different coupling coefficients $J$.
The short-circuit currents (see from the horizontal lines) increase with the increments of coupling intensities $J$, as shown by the curves in Fig.\ref{Fig.4}(c1), due to the same underlying physical significance as in Fig.\ref{Fig.2}. However, due to the correlation fluctuation intensities $\gamma_{22;22}$  of Site 2, some different physical properties are displayed. When compared to the results in Fig.\ref{Fig.2}, it is discovered that fluctuations play completely different roles in the EET process at the two different sites. Taking $J$=0.35 as an example, we can see that the descending short-circuit currents occur with $\gamma_{22;22}$ increments.

The underlying physical mechanisms for the different behavior of EET, we believe, are due to two factors bought by different correlated fluctuations at Sites 1 and 2, respectively. Site 2 is adjacent to the external terminal, as shown in Fig.\ref{Fig.1}. Even with a moderate fluctuation at Site 2, the EET is more susceptible to radiating to the ground state when it transports through Site 2 and completes the output photoelectric energy. Furthermore, compared to Site 1, some exciting energies on Site 2 emit out, and no more excitons are added to Site 2 as a result of solar photon non-absorption. This is another reason why EET decreases as the correlated fluctuations increase at Site 2. Fig.\ref{Fig.5} shows the output power due to different correlated fluctuations at Site 2 under the same parameter conditions as Fig.\ref{Fig.4}. Taking the curves corresponding to $J$=0.75, the descending peak powers due to increasing $\gamma_{22;22}$ demonstrate the consistent results obtained by Fig.\ref{Fig.4}.

\begin{figure}[htp]
\center
\includegraphics[width=0.45\columnwidth]{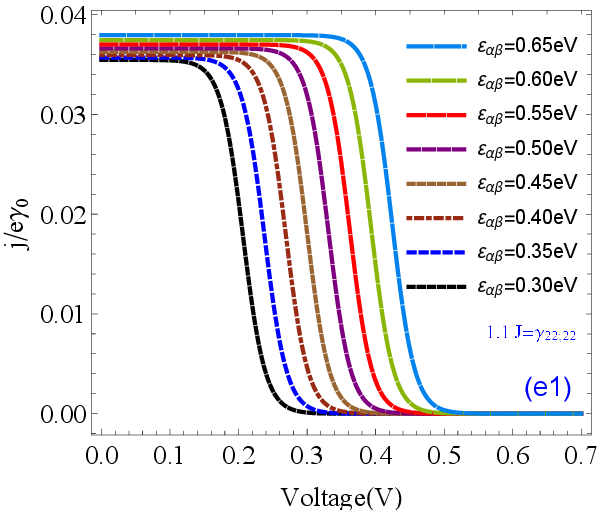}\includegraphics[width=0.45\columnwidth]{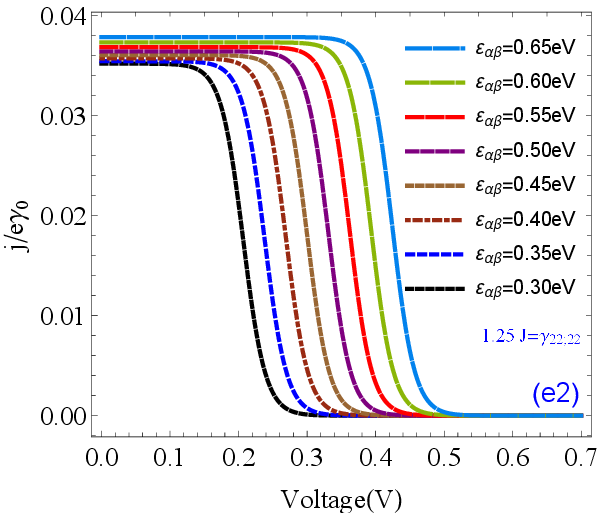}
\hspace{0in}%
\includegraphics[width=0.45\columnwidth]{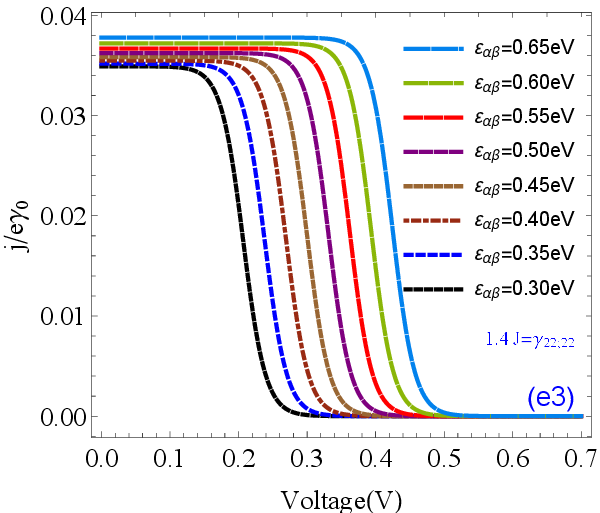}\includegraphics[width=0.45\columnwidth]{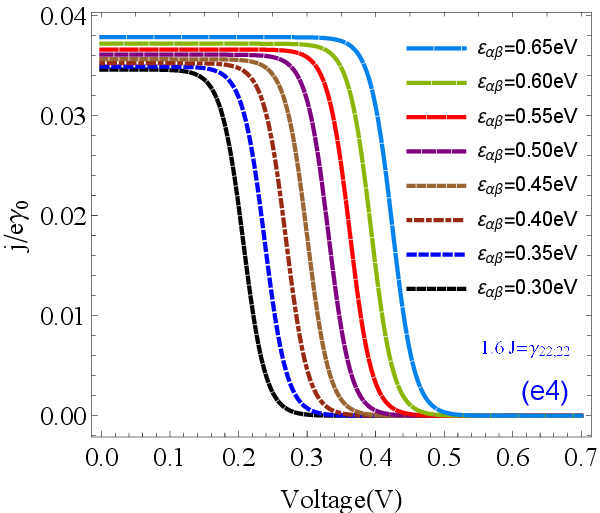}
\caption{(Color online) Current-voltage ($j$-$V$) characteristics assisted by the correlation fluctuation $\gamma_{22;22}$ of Site 2 with different gap energy differences $\varepsilon_{\alpha\beta}$. $\gamma_{11;12}$ - $\gamma_{22;12}$=3.2$\times10^{-6}$, $\gamma_{11;22}$=1.6$\gamma_{0}$, $\Gamma$=0.12$\gamma_{0}$, $\Gamma_{1c}$=0.09$\gamma_{0}$, $\Gamma_{2c}$=0.24$\gamma_{0}$, $J$=$0.4$, and all the other parameters are the same to those in Fig.\ref{Fig.4}.}
\label{Fig.6}
\end{figure}

\begin{figure}[htp]
\center
\includegraphics[width=0.45\columnwidth]{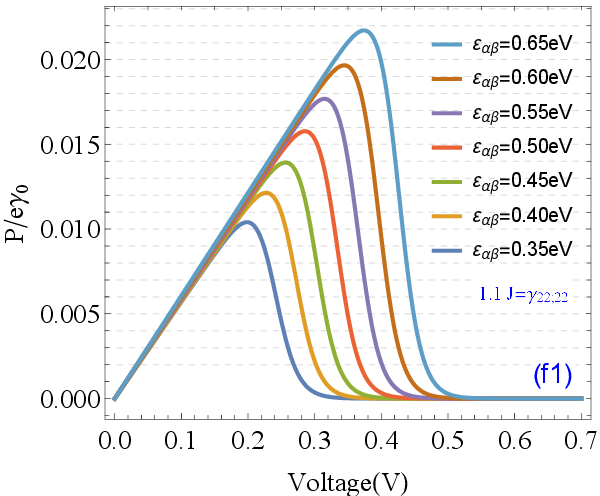}\includegraphics[width=0.45\columnwidth]{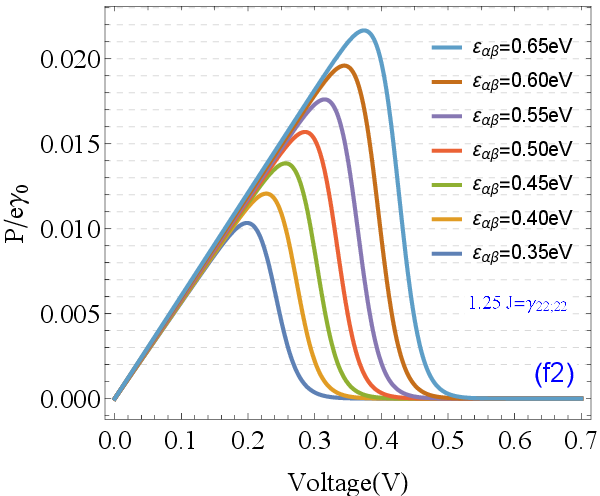}
\hspace{0in}%
\includegraphics[width=0.45\columnwidth]{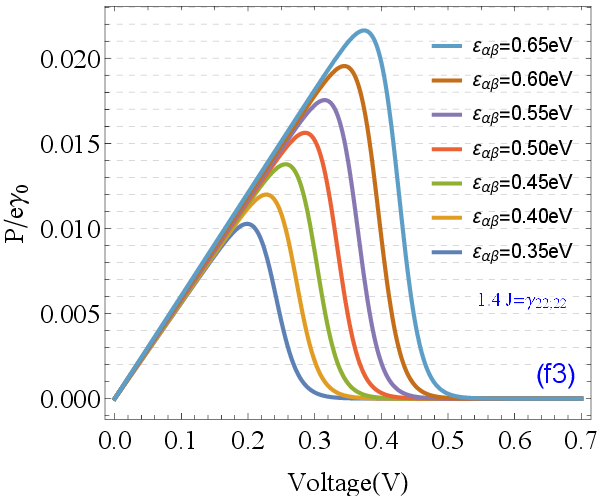}\includegraphics[width=0.45\columnwidth]{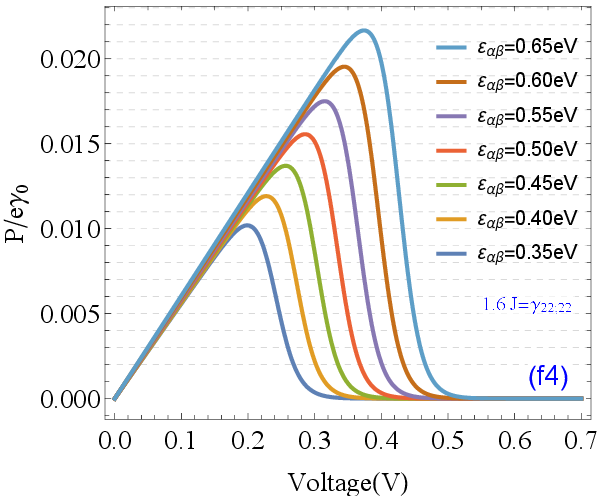}
\caption{(Color online) The output power $P$ versus voltage $V$ assisted by the correlation fluctuation $\gamma_{22;22}$ of Site 2 with different gap differences $\varepsilon_{\alpha\beta}$. All the other parameters are the same to those in Fig.\ref{Fig.6}.}
\label{Fig.7}
\end{figure}

The unabsorbed photons at Site 2 in this proposed model are adjacent to the ambient terminal, which influences the EET character. As a result, we believe that the gap energy $\varepsilon_{\alpha\beta}$ and correlated fluctuation at Site 2 may affect the EET in a completely different way. EET is plotted in Fig.\ref{Fig.6} and Fig.\ref{Fig.7} as a function of gap energy $\varepsilon_{\alpha\beta}$ and correlated fluctuation $\gamma_{22;22}$.
The curves in Fig.\ref{Fig.6} (e1) illustrate that the short-circuit currents ascend with the increments of gap energy differences $\varepsilon_{\alpha\beta}$. Meanwhile, the short-circuit currents increase with the increasing $\gamma_{22;22}$ from (e1) to (e3). However, we notice that a sharping decrease in short-circuit currents with $\gamma_{22;22}$=1.6$\gamma_{0}$, and a completely opposite behavior incurs due to the increasing $\varepsilon_{\alpha\beta}$ in the short-circuit currents in Fig.\ref{Fig.6} (e4). In other word, EET is more sensitive to $\gamma_{22;22}$ when compared these with the curves in Figs.\ref{Fig.4} and \ref{Fig.6}. Because of the increments in gap energy differences $\varepsilon_{\alpha\beta}$ in Fig.\ref{Fig.7}, the output power $P$ versus the output voltage $V$ further demonstrate the similar conclusions of $\gamma_{22;22}$ about EET.

In comparison to the results obtained by the correlated fluctuations at Site 1 in Fig.\ref{Fig.4} and Fig.\ref{Fig.5}, we believe the underlying physical mechanism for $\gamma_{22;22}$ in EET is different, as shown in Fig.\ref{Fig.6} and Fig.\ref{Fig.7}. When $\varepsilon_{\alpha\beta}$ has a larger value, the excitation energy has a greater transfer possibility at the localized charge-transfer state $|\alpha\rangle$, and the excitation energy is more likely to transport to the final charge-transfer state $|\beta\rangle$, which brings out the increasing short-circuit currents with the increments of $\varepsilon_{\alpha\beta}$ in Fig.\ref{Fig.6} (e1). Furthermore, when a minor correlated fluctuation occurs as a result of the interaction between Site 2 and the environment, the transfer rate accelerates and a higher EET occurs. The increasing short-circuit currents with $\gamma_{22;22}$ increments demonstrate these conclusions from Fig.\ref{Fig.6} (e1) to (e3). When a moderately larger correlated fluctuation occurs at Site 2, however, the transfer process of partial excitation energy at the localized charge-transfer state $|\alpha\rangle$ is disrupted and radiates out, resulting in less excitation energy being transported to the final charge-transfer state $|\beta\rangle$. Furthermore, the transfer rate is delayed as a result of the larger gap difference $\varepsilon_{\alpha\beta}$. The longer the transfer process, the greater the gap difference $\varepsilon_{\alpha\beta}$. As a result of the lengthy transfer process, the excitation energy arrives at the final charge-transfer state $|\beta\rangle$ with a lower EET rate to the final charge-transfer state $|\beta\rangle$. The damping shirt-circuit currents increased with increasing $\varepsilon_{\alpha\beta}$, confirming the above analysis in Fig.\ref{Fig.6} (e4). Under the same conditions as in Fig.\ref{Fig.7}, the output power $P$ versus voltage $V$ demonstrates a physical regime similar to those in Fig.\ref{Fig.6}.

Before concluding this section, we'd like to point out that some research has shown that quantum coherence in photosynthetic light-harvesting complexes can boost the EET. Despite the fact that the destructiveness of quantum coherence has been studied due to the correlated fluctuations at site-energy, the quantitative relationship and physical mechanism are not carried out in the pigment molecular chains. The significance of this work differs from that of Ref.\cite{Abramavicius2011}, which incorporates uncorrelated and correlated fluctuations into exciton dynamics and provides insight into two-dimensional optical spectroscopy of photosynthetic complexes, with the results indicating that correlated fluctuations do not affect single-exciton dynamics and have a nonlocal contribution to spectral broadening. We argue that understanding the relationship between correlated fluctuation and quantum coherence reveals the near-unity quantum efficiency of EET among pigments in a noisy environment and inspires one to understand its mechanism in order to elucidate design principles for bioinspired artificial light harvesting. This is a working domain and direction that deserves some future attention.

\section{Remarks and conclusions}

In conclusion, we distinguished the functions of correlated fluctuations in EET in two adjacent pigments in the noisy environment of pigment-protein complexes in this paper. In this proposed scheme, unlike previous work on the EET in pigment chains with an arbitrary number of sites, the influence of correlated fluctuations at two adjacent pigments was discussed. Because of their different underlying physical mechanisms, the pigment molecule acting as the charge-transfer excited state, i.e., Site 2, is more sensitive to correlated fluctuations than the pigment molecule acting as the photon-absorbing excited state, i.e., Site 1. As a result, different pigments in photosynthetic light-harvesting complexes play different roles in EET. The schemes in which pigments were assumed to be arranged in a similar pattern and respond to a uniform environment in pigment-protein complexes may have overlooked some other underlying physical significance.

\section*{Acknowledgments}

This work is supported by the National Natural Science Foundation of China ( Grant Nos. 62065009 and 61565008 ), Yunnan Fundamental Research Projects, China( Grant No. 2016FB009 ).

\bibliography{reference}
\bibliographystyle{unsrt}
\end{document}